\documentclass[aps, prd, reprint, notitlepage, a4paper, floats, amsmath, amssymb, amsfonts,superscriptaddress,showpacs, showkeys,nofootinbib,longbibliography]{revtex4-2}

\pdfoutput=1

\usepackage{graphicx}
\usepackage[dvipsnames]{xcolor}
\usepackage[normalem]{ulem}
\usepackage{booktabs}
\definecolor{dodgerblue}{HTML}{1E90FF}
\definecolor{viennared}{HTML}{DA0A14}
\usepackage[english]{babel}
\allowdisplaybreaks[1]
\usepackage[utf8]{inputenc}
\usepackage{lmodern}
\usepackage{booktabs}
\usepackage{diagbox}
\usepackage{dcolumn}
\usepackage{natbib}
\usepackage{mathrsfs}
\usepackage{amssymb}
\usepackage{physics}
	\newcolumntype{.}{D{.}{.}{13}}
	\newcolumntype{d}[1]{D{.}{.}{#1}}

\usepackage[colorlinks=true,
            citecolor=blue,
            linkcolor=blue]{hyperref}

\usepackage{cleveref}

\allowdisplaybreaks
\graphicspath{{fig/}}

\newcommand{\UMONSgrav}{\affiliation{Physique de l'Univers, Champs et Gravitation, 
Universit\'e de Mons -- UMONS,\\Place du Parc 20, 7000 Mons, Belgium}}
\newcommand{\UMONSnuc}{\affiliation{Physique nucléaire et subnucléaire, 
Universit\'e de Mons -- UMONS,\\Place du Parc 20, 7000 Mons, Belgium}}

\newcommand{\beq}{\begin{equation}}
\newcommand{\eeq}{\end{equation}}
\newcommand{\bea}{\begin{eqnarray}}
\newcommand{\eea}{\end{eqnarray}}
\newcommand{\bit}{\begin{itemize}}
\newcommand{\eit}{\end{itemize}}
\newcommand{\ben}{\begin{enumerate}}
\newcommand{\een}{\end{enumerate}}

\begin{document}

\title{
	A Plunge into the Chasm: Surviving Tidal Effects in Kerr Spacetime 
}
\author{Guillaume Lhost}
\email[E-mail: ]{guillaume.lhost@umons.ac.be}
\thanks{ORCiD: 0009-0007-6024-6896}
\UMONSgrav
\author{Ornella Ruta}
\email[E-mail: ]{ornella.ruta@student.umons.ac.be}
\UMONSgrav
\author{Claude Semay}
\email[E-mail: ]{claude.semay@umons.ac.be}
\thanks{ORCiD: 0000-0001-6841-9850}
\UMONSnuc

\date{\today}

\begin{abstract}
We investigate the fate of an observer falling towards a Kerr black hole. 
The tidal forces are computed for arbitrary trajectories of an observer, and we specify them along the polar axis in order to remain as far as possible from the ring-shaped singularity.
Our analysis shows that an observer is not tidally disrupted during the fall provided that the black hole mass exceeds a critical value, which depends on its spin. 
In practice, any supermassive black hole represents a suitable candidate to allow an observer to traverse the black hole without severe deformation.
In contrast, stellar-mass rotating black holes do not satisfy the mass condition and are expected to subject the observer to extreme tidal forces leading to its destruction during the plunge.
\end{abstract}

\maketitle


\section{Introduction}

It is well known that an object in free fall towards a static black hole will undergo spaghettification \cite{Misner:1973prb}, which inevitably leads to its destruction before it even reaches the singularity. However, the situation is quite different for a Kerr black hole, where tidal effects differ due to the ring-shaped nature of the singularity \cite{LimaJunior:2020fhs}. 
\\
The purpose of this paper is to examine to what extent an observer in free fall in the vicinity of a Kerr black hole can survive and reach the center of the ring-shaped singularity unharmed. 
The first step is to compute the tidal effects on a small body plunging into a Kerr black.
The body can follow a geodesic trajectory confined on a fixed polar plane, and the proper time till the annular singularity is finite.
Then, we compute tidal forces acting on the body along its trajectory, and deduce the stress acting on it. 
We then restrict attention to motion along the symmetry axis, where tidal effects are minimal as the body is the furthest from the spacetime singularity.
We find that the tidal forces and pressure on this axis are maximal at a given radial distance from the singularity. We show that the characteristics of a Kerr black hole can allow survival of the falling observer, which will eventually reach the singularity.

No need of super strength or ultra-strong armor to pass through it, provided the black hole is massive enough and spins fast enough. Any supermassive black hole in galactic cores probably meets the requirements. 
All of our results are analytical. 
\\
The Kerr spacetime and the plunge trajectories towards the black hole are described in Sect.~\ref{sec:Kerr}. Sect.~\ref{sec:Tidal} is devoted to the computation of the tidal accelerations. The maximum tidal stress acting on a free-falling body and the black hole critical mass to survive the plunge are computed in Sect.~\ref{sec:Max}. A ``sci-fi" scenario for the destiny of an observer reaching the center of the singularity is proposed in Sect.~\ref{sec:SciFi}. Final remarks are given in Sect.~\ref{sec:Conclu}. 

\section{Kerr spacetime and plunge trajectories}
\label{sec:Kerr}

\subsection{Kerr solution}
We work in geometric units $G = c =1.$
The background spacetime is described by the Kerr metric 
$g_{\mu\nu}$ of an isolated black hole. 
We work in Boyer-Lindquist (BL) coordinates $(t, x^i)=(t, r, \theta, \phi)$, where $t$ and $r$ have the dimension of mass. In BL coordinates, the line element built with the Kerr metric reads \cite{Misner:1973prb}
\begin{align} \label{eq:Kerr metric}
    ds^2 = -&\left(1 - \frac{2 r}{M\,\rho^2(r,\theta)}\right)dt^2 - 2\frac{2  a \, r\sin^2 \theta}{M \,\rho^2(r,\theta)} dt d\phi\nonumber\\
    &+M^2 \frac{\Sigma(r,\theta)}{\rho^2(r,\theta)}\sin^2 \theta d\phi^2 + \frac{\rho^2(r,\theta)}{\Delta(r)} dr^2 \nonumber\\ 
    &+ M^2 \rho^2(r,\theta) d\theta^2,
\end{align}
where we have defined the dimensionless functions
\begin{equation}
\begin{aligned}
    \rho^2(r,\theta) &:= \frac{1}{M^2} \left( r^2 + a^2 \cos^2 \theta\right),\\
    \Delta(r) &:= \frac{1}{M^2} \left( r^2 - 2 M r + a^2\right),\\
    \Sigma(r,\theta) &:= \frac{1}{M^4} \left( \big(r^2 + a^2\big)^2 - M^2 a^2 \Delta(r) \sin^2 \theta \right).
\end{aligned}
\end{equation}
The parameter $M$ designates the mass of the black hole while the parameter $a$ ($[a] = \textrm{M}$) is the spin of the black hole. In the limit $a \rightarrow 0$, the metric $\eqref{eq:Kerr metric}$ reduces to the Schwarschild metric in BL coordinates.
The Kerr metric admits two coordinate singularities $r_+$ and $r_-$ given by
\begin{equation} \label{eq: coordinate singularities}
    r_\pm = M  \pm \sqrt{M^2- a^2}.
\end{equation}
The Penrose censorship ensures that these singularities must be real. 
This imposes a constraint on the spin of the black hole which must satisfy $|a| \leq M$.
On the other hand, the spacetime singularity is found in $r= 0$ when $\theta = \frac{\pi}{2}$. 
This singularity has the geometric shape of a ring lying on the equatorial plane with radius $|a|$.

\subsection{Geodesics on a plane}
We investigate the possibility for a test body to fall radially towards the singularity while remaining confined to a constant polar angle, $\theta = \theta_0$.
In general, when the constants of motion are independent, the motion in the polar direction is not fixed: the orbital plane precesses, and the particle oscillates between two extremal values of the polar angle.
In the present analysis, such behavior must be avoided, as reaching $\theta = \tfrac{\pi}{2}$ would result in extreme tidal forces in the vicinity of $r = 0$.
\\
The worldline of a test body in Kerr spacetime is given by 
\begin{equation}
    z^\mu_p(\tau) = (t_p(\tau),r_p(\tau), \theta_p(\tau),\phi_p(\tau)),
\end{equation}
and the 4-velocity is given
\begin{equation}
    u_p^\mu(\tau) = (\dot{t}_p(\tau),\dot{r}_p(\tau), \dot{\theta}_p(\tau),\dot{\phi}_p(\tau)),
\end{equation}
where a dot corresponds to a derivative with respect to the proper time $\tau$. In the following, we will sometimes make use of the Mino time $\lambda$ parametrization which is relevant for decoupling the polar and radial motion of the test body. 
This Mino time is defined by
\begin{equation} \label{eq: Mino}
    \frac{d\lambda}{d\tau} = \frac{1}{\rho^2(r,\theta)}.
\end{equation}
A geodesic trajectory is given by the conservation law of four constants of motion. The isometries of the Kerr spacetime (time translation and axial symmetries) immediately define the energy and the axial angular momentum per unit mass of the test body
\begin{equation}
    E = - u_t, \quad L = u_\phi.
\end{equation}
We have $[E] = 1$ and $[L]=\textrm{M}$.
The Carter constant is also conserved; it arises from the Killing tensor of Kerr. 
See also Carter and Schmidt \cite{Carter:1968rr,Schmidt:2002qk} for description of the Carter constant.
We define the Carter constant (per unit mass square of the test body) with
\begin{equation}
    Q =u_\theta^2
+ a^2 \cos^2\theta
+ \left( a E \sin\theta - \frac{L}{\sin\theta} \right)^2
- (L - a E)^2
\end{equation}
It is defined such that $Q=0$ in the equatorial plane.
Together with the mass-shell constraint of a massive body on a time-like geodesic $g_{\alpha \beta }u^\alpha u^\beta = -1$, the conservation laws of energy, angular momentum and Carter constant are equivalent to the first order differential equations 
\begin{subequations}
\begin{align}
    \left( \frac{d r_p}{d\lambda}\right)^2 &= \mathcal{R}(r_p),
    \\
    \left( \frac{d \theta_p}{d\lambda}\right)^2 &= \Theta(\theta_p),
    \\
    \frac{dt_p}{d\lambda} &= \mathcal{T}_r(r_p) + \mathcal{T}_\theta (\theta_p),
    \\
    \frac{d\phi_p}{d\lambda} &= \Phi_r(r_p) + \Phi_\theta (\theta_p)
\end{align}
\end{subequations}
where
\begin{subequations}
\begin{align}
\mathcal{R}(r) &:= \frac{1}{M^4} \left[ \left( (r^2 + a^2)E - aL \right)^2 \right]
\\
&- \frac{\Delta(r)}{M^2} \left[ r^2 + (aE - L)^2 + Q \right], \nonumber \\[0.5em]
\Theta(\theta) &:= \frac{1}{M^4} \left[ Q - \cos^2\theta \left( a^2(1 - E^2) + \frac{L^2}{\sin^2 \theta}  \right) \right], 
\\[0.5em]
\Phi_r(r) &:= \frac{a}{M^4 \,\Delta(r)} \left[ E(r^2 + a^2) - aL \right] - \frac{aE}{M^2}, 
\\[0.5em]
\Phi_\theta(\theta) &:= \frac{L}{M^2 \sin^2\theta}, 
\\[0.5em]
\mathcal{T}_r(r) &:= \frac{r^2 + a^2}{M^4 \,\Delta(r)} \left[ E(r^2 + a^2) - aL \right], \\[0.5em]
\mathcal{T}_\theta(\theta) &:= -\frac{a}{M^2} \left( aE \sin^2\theta - L \right).
\end{align}
\end{subequations}
In these equations, we took advantage of the Mino Time parametrization to decouple the $r$ and $\theta$ dependence in the equations of motion. 
This allows to define two distinct effective radial and polar potentials $\mathcal{R}$(r) and $\Theta(\theta)$.
\\
We focus on trajectories corresponding to a plunge confined to a fixed polar plane, starting from a large radial distance $r_0 \gg r_+$ with energy $E \geq 1$.
A geodesic can remain at a constant polar angle $\theta_0 \neq \frac{\pi}{2}$ only if the polar potential satisfies
$   
\Theta(\theta_0)=0=\Theta'(\theta_0).
$
These conditions impose constraints on the angular momentum and the
Carter constant, which take the form
\begin{equation}
\begin{aligned} 
    L &\equiv L_0 := \pm  \,a \sqrt{E^2 - 1}\,\sin^2(\theta_0), \label{eq: polar plane constraints} \\
    Q &\equiv Q_0 := -a^2 (E^2 - 1)\, \cos^4(\theta_0).
\end{aligned}
\end{equation}
In the limit $E = 1$, the constraint reduces to $L = Q = 0$. Note that the particular case of $\theta_0 = \frac{\pi}{2}$ immediately ensures that the polar potential and its derivative vanishes, whatever the angular momentum.
A perturbation of the angular momentum, $L_0 \rightarrow L_0 + \delta L$,
must be accompanied by a corresponding shift $Q_0 \rightarrow Q_0 + \delta Q$ so that the polar-plane condition \eqref{eq: polar plane constraints} remains satisfied for a slightly modified angle $\theta_0 \rightarrow \theta'_0$.
If this compensation does not occur, the plunging test body will begin to
oscillate between two polar regions, repeatedly switching from the north pole
to the south pole and vice versa, instead of remaining on a fixed polar plane.
Such shifts may be interpreted as the need for the infalling
object to use a motor or other stabilizing mechanism to counter external
perturbations that could otherwise alter its constants of motion.
\\
Assume that the conditions \eqref{eq: polar plane constraints} hold for some
angle $\theta_0 \neq \frac{\pi}{2}$, and suppose that an external perturbation
slightly modifies the polar angle so that
$\theta_0 \rightarrow \theta_0' = \theta_0 + \delta\theta$.
In this case, the deviation $\delta\theta$ satisfies the equation $ \frac{d^2}{d\lambda^2}\,\delta\theta = - \omega^2 \delta\theta$ ,
with
$\omega^2 = \frac{4 a^2 (E^2 - 1)\cos^2(\theta_0)}{M^4}.$ 
This positive $\omega^2$ ensures that a free-fall with a fixed polar angle is a stable trajectory. 
The frequency $\omega$ vanishes only in the limit $E \rightarrow 1$, and it
reaches its maximum when the plunge occurs near the north or south pole.
\\
Now we compute the total amount of proper time for a plunge starting from $r = r_0$ down to the singularity at $r = 0$, while keeping the polar angle fixed. 
We assume that the momentum and the Carter constant satisfy the constraints given in Eq.~\eqref{eq: polar plane constraints}. 
The proper time $\Delta \tau$ of the free fall is given by
\begin{equation} \label{eq: delta tau}
    \Delta \tau(\theta_0) = \int_{r_0}^{0} -\frac{\rho^2(r',\theta_0)}{\sqrt{\mathcal{R}(r')}} \, dr' \, .
\end{equation}
This integral can be evaluated in terms of elliptic integrals or hypergeometric functions, as shown in \cite{Dyson:2023fws}. 
We have numerically verified that the integral \eqref{eq: delta tau} is finite for any value of $\theta_0$, for any $E > 1$, and for arbitrary Kerr parameters $\{a, M\}$. 
In particular, in the limit $E \to 1$ and for $\theta_0$ equal to $0$ (i.e.\ for an observer starting with a non-relativistic velocity near the polar axis), we obtain
\begin{equation}
    \Delta \tau(0)\Big|_{E=1} 
    = \sqrt{\frac{2 r_0}{M}} \, |a| \, 
    {}_2F_1\!\left(-\frac{1}{2},\,\frac{1}{4};\,\frac{5}{4};\,-\frac{r_0^2}{a^2}\right),
\end{equation}
where ${}_2F_1$ denotes the Gauss hypergeometric function. 
This result shows that the total proper time along the polar axis is always finite. Therefore, a plunge along a fixed plane is physically possible, and an infalling observer reaches the singularity in a finite amount of proper time.

\section{Tidal accelerations in Kerr spacetime}
\label{sec:Tidal}

Consider two nearby points in Kerr spacetime. 
Treated as test particles, these elements follow two distinct geodesics. 
Let $\xi^\mu$ denotes the separation vector connecting these neighboring geodesics. 
The relative acceleration of this vector is governed by the geodesic deviation equation:
\begin{equation}
    \frac{D^2 \xi^\mu}{d\tau^2} = - R^\mu_{\ \nu\alpha\beta}\, u^\nu \, \xi^\alpha \, u^\beta \,,
\end{equation}
where $u^\mu$ is the four-velocity of the reference geodesic and $R^\mu_{\ \nu\alpha\beta}$ is the Riemann curvature tensor. 
If these nearby points represent adjacent, fixed elements within the body of an observer falling into a black hole, then the magnitude of $\frac{D^2 \xi^\mu}{d\tau^2}$ can be interpreted as the tidal acceleration required to keep these elements at a fixed separation.
\\
It is convenient to choose a local frame in which the spatial velocity of an infalling object vanishes. 
For that purpose, we work in an orthonormal frame associated with the Kerr spacetime. 
Namely, we introduce the vierbein fields $e_{\mu}^{\hat{a}}$ and their inverse $e^{\mu}_{\hat{a}}$, defined such that
\begin{align}\label{eq:orthoframe}
e^{\mu}_{\hat{a}} \, g_{\mu\nu} \, e^{\nu}_{\hat{b}} &= \eta_{\hat{a}\hat{b}}, 
\qquad
e_{\mu}^{\hat{a}} \, e^{\mu}_{\hat{b}} = \delta^{\hat{a}}_{\hat{b}},
\end{align}
where $\eta_{\hat{a}\hat{b}} = \mathrm{diag}(-1, +1, +1, +1)$ is the Minkowski metric. 
Latin indices with a hat refer to the local Lorentz (non-coordinate) frame, while Greek indices denote spacetime coordinates.
\\
The tetrad provides a local orthonormal basis at each point of the spacetime. 
It allows to project any tensorial quantity onto a locally inertial frame. 
In particular, for a given vector $V^\mu$, its components in the local Lorentz frame are given by 
$V^{\hat{a}} = e^{\hat{a}}_\mu \, V^\mu$.
\\
In the following, we choose the tetrad \cite{Mino:1995fm}
\begin{equation}
\begin{aligned}
e_t^{\hat{0}} &= \sqrt{\frac{\Delta(r)}{\rho^2(r,\theta)}}, 
\qquad \qquad \quad 
e_t^{\hat{3}} = -\frac{a \sin\theta}{M\,\rho(r,\theta)}, \\[0.5em]
e_r^{\hat{1}} &= \sqrt{\frac{\rho^2(r,\theta)}{\Delta(r)}}, 
\qquad \qquad \quad 
e_\theta^{\hat{2}} = M\,\rho(r,\theta), \\[0.5em]
e_\phi^{\hat{0}} &= -a \sin^2\theta \,\sqrt{\frac{\Delta(r)}{\rho^2(r,\theta)}},
\quad 
e_\phi^{\hat{3}} = \frac{r^2 + a^2}{M\,\rho(r,\theta)} \sin\theta.
\end{aligned}
\end{equation}
By construction, this tetrad satisfies the properties of Eq.~\eqref{eq:orthoframe}.
\\
In this Lorentz frame, the four-velocity of a test particle is
$u^{\hat{a}} = (1,0,0,0).$
This reduces the geodesic deviation to
\begin{equation} \label{eq:tidal accel}
\frac{d^2 \xi^{\hat{a}}}{d\tau^2} = - R^{\hat{a}}_{\ \hat{0}\hat{0}\hat{d}} \, \xi^{\hat{d}},
\end{equation}
The components $R_{\hat{a}\hat{0}\hat{0}\hat{d}}$ admit a direct physical interpretation: they describe the tidal accelerations between two freely falling bodies. 
Equivalently, they can be understood as the force per unit mass required to keep the distance between two nearby freely falling particles constant.
Note that, as $\xi^{\hat{a}}$ represents the separation between two nearby objects, tidal forces can be compared to those exerted by a string joining the two objects.
\\
According to Eq.~\eqref{eq:tidal accel}, the tidal acceleration of the separation vector between two adjacent geodesics depends on the coordinates $r_p$ and $\theta_p $, which localize the body.
For any $\theta_p$, the tidal accelerations are given by
\begin{equation}
    \begin{aligned}
        \label{eq: tidal accel full}
\frac{d^2 \xi^{\hat{0}}}{d\tau^2} &= 0, 
\\
\frac{d^2 \xi^{\hat{1}}}{d\tau^2} &= \frac{8 M \,r_p (2r_p^2 - 3 a^2-3 a^2 \cos(2 \theta_p))}{(a^2+2 r_p^2+a^2 \cos(2 \theta_p))^3} \xi^{\hat{1}},
\\
\frac{d^2 \xi^{\hat{b}}}{d\tau^2} &= - \frac{4 M \,r_p (2 r_p^2 - 3 a^2-3 a^2 \cos(2 \theta_p))}{(a^2+2 r_p^2+a^2 \cos(2 \theta_p))^3} \xi^{\hat{b}},
    \end{aligned}
\end{equation}
where $\hat{b} \in \{\hat{2},\hat{3}\}$.
The limit $\theta_p \to 0$ is straightforward.
These accelerations are consistent with those derived in \cite{LimaJunior:2020fhs} for a particle moving along the rotation axis of a Kerr black hole. 
Moreover, taking the direct limit $a=0$ reproduces the tidal forces experienced by a freely falling observer in a Schwarzschild spacetime.
Equations~\eqref{eq: tidal accel full} are symmetric under the transformation $\theta_p \to \pi - \theta_p$ and depend only on $a^2$, showing that there is no distinction between plunging through the north or south pole; the infalling observer experiences the same stretching in either case.
Note that $\frac{d^2 \xi^{\hat{1}}}{d\tau^2}$ corresponds to the tidal force acting in the radial direction, towards the center of the annular singularity, whereas $\frac{d^2 \xi^{\hat{b}}}{d\tau^2}$ represents the tidal forces in the polar and azimuthal directions, i.e. those perpendicular to the radial direction.

\section{Maximal tidal stress and critical mass}
\label{sec:Max}

Let us consider a human body of mass $m$ freely falling towards $ r = 0$  while remaining on the symmetry axis.
We fix $ \theta_p = 0 $. 
The human body is modeled as a set of points forming a homogeneous parallelepiped of length $ L$ and transverse width $\ell$.
Its center of mass (CM) is at the radial coordinate $r_p$.
We introduce a Cartesian coordinate system $ \{x, y, z\}$, with basis vectors $\{\vec{1}_x, \vec{1}_y, \vec{1}_z\}$, whose origin is located at the CM of the parallelepiped.
The vector $\vec{1}_x$ is aligned with $e^{\hat{1}}$, while $\vec{1}_y$ and $\vec{1}_z$ are aligned with $e^{\hat{2}}$ and $ e^{\hat{3}}$, respectively.
Each volume element of the parallelepiped is connected to the CM by a separation vector 
$
\vec{\xi}_i(x,y,z) = x_i \vec{1}_x + y_i \vec{1}_y + z_i \vec{1}_z 
$
which is subjected to an infinitesimal force
$
d\vec{F}_i = \ddot{\vec{\xi}}_i \, dm ,
$
where $dm = \frac{m}{L\,\ell^2} \, dx\,dy\,dz $ is the mass element, and $\ddot{\vec{\xi}}_i$ denotes the accelerations computed from Eq.~\eqref{eq: tidal accel full} for a fixed value of $r_p$ and with $\theta_p =0$ .
\\
We are interested in computing the tidal stress tensor to which the observer is subjected. 
By symmetry of the body, this tensor is diagonal in the principal axes $\{x,y,z\}$.
To this end, we first consider the $yz$-plane passing through the CM. 
The longitudinal tidal force is defined as the sum of all force elements contributing to the stress perpendicular to this plane, namely along the $x$-direction. 
It is given by
\begin{equation}
    F^x = \int_{yz}\int_0^{L/2} dx \,dy\,dz \, \ddot{\xi}^z \,\frac{m}{L \ell^2}
    = \frac{L \,m\,M\, r_p(r_p^2-3a^2)}{4(a^2+r_p^2)^3}.
\end{equation}
Similarly, the transverse tidal forces across the $yz$ and $xz$ planes are
\begin{align}
    F^y = F^z = -\frac{\ell \,m\,M\, r_p(r_p^2-3a^2)}{8(a^2+r_p^2)^3}.
\end{align}
These components can be interpreted as the magnitude of the internal forces that must be sustained by the body in order to preserve its internal structure against tidal deformation. 
The tidal field tends to stretch the body along the $x$-direction and compress it along the transverse directions.
The tidal stress tensor $T^{ij}$ is then obtained by dividing the opposite of each force component by the corresponding surface area perpendicular to the direction of the force. This yields
\begin{equation} \label{eq: stress tensor}
    T^{ij} = \frac{m\,M\, r_p(r_p^2-3a^2)}{8(a^2+r_p^2)^3}
    \begin{pmatrix}
        \frac{-2 L}{\ell^2} & 0 & 0 \\
        0 & \frac{1}{L} & 0 \\
        0 & 0 & \frac{1}{L}
    \end{pmatrix}.
\end{equation}
We note that the stress vanishes at $r_p=\sqrt{3}\,a$, below which it changes sign.
\\
Let $P_L$ and $P_T$ denote the maximum longitudinal and transverse pressures that a human body can tolerate before being torn apart or fatally compressed. 
We aim to determine whether there exist values of $M$ and $a$ such that the tidal effects remain below these critical thresholds for all $r_p$, down to the singularity.
To this end, it is suitable to introduce the dimensionless parameters $(\alpha,\beta, n)$ defined by
\begin{equation}
\begin{aligned}
    a &= \alpha \, r_g,
    \\
    r_p &= \beta \, r_g,
    \\
    M &= n \, M_\odot,
\end{aligned}
\end{equation}
where $M_{\odot}$ is the solar mass and 
\begin{equation}
    r_g  = n\,M_\odot,
\end{equation}
is half of the Schwarzschild radius.
By introducing these parameters, the mass dimension of any parameter characterizing the black hole or the trajectory can be explicitly read through $M_\odot$.
We then examine the radial coordinate at which the absolute value of the stress tensor components is maximized. 
For each component, the global maximum is reached for $\beta < \sqrt{3}\,\alpha$. It corresponds to
\begin{equation} \label{eq: radius with max stress}
    \beta_{max} = \sqrt{\alpha^2\left(3 - 2\sqrt{2}\right)} .
\end{equation}
At this location, the longitudinal and transverse stresses scale as $n^{-2}$, indicating that a larger black hole mass results in weaker tidal effects on the body.
This scaling is identical to that of the density of the rotating black hole, assuming that this latter is modeled as a spherical object of mass $M$ with a radius $r_+$ given in Eq.~\eqref{eq: coordinate singularities}.
We investigate whether the stress can become so large that it will overtake $P_L$ or $P_T$.
We therefore define the following  polynomials 
\begin{equation}
\begin{aligned}
    \mathscr{P}_L(\beta) &= \left( P_L - |T^{xx}| \right)\, 4l^2 (\alpha^2+\beta^2)^3 \, r_g^6,
    \\
     \mathscr{P}_T (\beta) &= \left( P_T - |T^{yy}| \right)\, 8L(\alpha^2+\beta^2)^3 \, r_g^6. 
\end{aligned}
\end{equation}
If, for all values of $\beta$, both polynomials are strictly positive, then the infalling body remains intact; otherwise, it is destroyed.
The most critical location corresponds to $\beta = \beta_{\mathrm{max}}$. 
We then determine the critical values of $n$ from the conditions
\begin{align}
    \mathscr{P}_L(\beta_{\mathrm{max}})=0
    \quad \text{and} \quad
    \mathscr{P}_T(\beta_{\mathrm{max}})=0,
\end{align}
which respectively yield
\begin{equation}
\begin{aligned}
    n_L &= \frac{\left( 2+ \sqrt{2}\right)\sqrt{L \,m}}{8 \ell \, M_{\odot} |\alpha|^{3/2} \sqrt{P_L}},
    \\
    n_T &= \frac{\left( 1+ \sqrt{2}\right)\sqrt{m}}{8 \, M_{\odot} |\alpha|^{3/2} \sqrt{L \,P_T}}.
\end{aligned}
\end{equation}
We now have a criterion for the survival of an observer falling into a Kerr black hole of dimensionless spin $\alpha$.
Assuming $P_L = P_T$, we conclude that the observer is not tidally broken during the infall provided that the black hole mass exceeds a critical value. 
Restoring SI units, this critical mass is given by
\begin{equation} \label{critical mass formula}
    M_{\mathrm{crit}} = \frac{c^3}{G} \frac{\left( 2+ \sqrt{2}\right)}{8 \ell \, |\alpha|^{3/2} } \sqrt{\frac{L \,m}{P_L}}.
\end{equation}
According to this result, any black hole with mass $M > M_{\mathrm{crit}}$ can, be traversed without tidal disruption along the polar axis.
We have thus identified the minimum mass a Kerr black hole must possess for an observer to survive the plunge down to the singularity. 
The most dangerous region occurs at $r_p = \beta_{\mathrm{max}}\,r_g$, where tidal stresses reach their maximum.
For a given dimensionless spin $\alpha$, any mass larger than $M_{\mathrm{crit}}$ ensures survival through this critical region. In particular, for an extremal black hole ($\alpha = 1$), the critical mass is approximately $M_{\mathrm{crit}} \simeq 8.66 \times 10^4\sqrt{\frac{L \,m}{\ell^2 P_L}}
\,M_\odot$.
As $\alpha \to 0$ (Schwarzschild limit, for which $\beta_{\mathrm{max}} = 0$), the critical mass diverges, meaning that an infinitely massive black hole would be required to avoid tidal disruption at the singularity. This highlights that, in the non-rotating case, tidal forces inevitably become destructive.

\subsection{Explicit computation for Sagittarius A*}

We use the results of \cite{Daly:2023axh}, which estimate the dimensionless spin parameter of the Sagittarius $A^\star$ black hole to be 
$
\alpha_\star = 0.90 \pm 0.06 .
$
Its mass is given by
$
M_\star = (4.152 \pm 0.014)\times 10^{6}\, M_{\odot}.
$
For definiteness, we consider an observer of mass $m = 75\,\mathrm{kg}$, height $1.70\,\mathrm{m}$, and width $0.40\,\mathrm{m}$. 
We further assume that the maximal longitudinal stress sustainable by the human body is $P_L = 100\,\mathrm{atm}$ \cite{Misner:1973prb}.
\\
Using Eq.~\eqref{critical mass formula}, we find that the critical black hole mass corresponding to the spin of Sagittarius $A^\star$ is
\begin{equation}    M^\star_{\mathrm{crit}} \simeq 900\, M_\odot .
\end{equation}
This value is more than three order of magnitude smaller than the mass of Sagittarius $A^\star$.
We conclude that an observer plunging towards the center of the ring-shaped singularity would not experience fatal tidal stresses.
Actually, using Eqs.~\eqref{eq: stress tensor} and \eqref{eq: radius with max stress}, we find that the maximum stress experienced by the body during the plunge is smaller than
$
0.5\,\mathrm{Pa}.
$
Such a value is negligible compared to the mechanical stress sustainable by the human body. 
Therefore, tidal effects would have essentially no impact on an observer plunging into a supermassive black hole such as Sagittarius $A^\star$.
\\
Note that, for this mass and spin, the radius of the singularity is approximately
$ 5\times10^{6}\,\mathrm{km},$
which is many orders of magnitude larger than the size of the human body. 
This justifies the use of the geodesic deviation equation at first order in the separation distance between different points of the body.
\\
We conjecture that any supermassive black hole with a mass exceeding $10^6$ solar masses can host an observer without causing major tidal damage.
Indeed, according to the critical mass formula \eqref{critical mass formula}, and using the parameters of the observer defined earlier in this section, such black holes require a dimensionless spin parameter larger than $8.39 \times 10^{-3}$.
This threshold is very small compared to the spin measurements of supermassive black holes summarized in Table I of \cite{Reynolds:2020jwt}.
However, an astrophysical stellar-mass black hole, such as the remnant of the recently detected  gravitational wave GW250114 event \cite{LIGOScientific:2025rid}, is not a suitable candidate to host the observer. 
In particular, the remnant spin has been measured to be $\alpha =0.68 \pm 0.01$, while its mass is $(62.7 \pm 1.0) \,M_{\odot}$. This lies below the threshold set by the critical mass formula.

\section{Sci-fi scenario}
\label{sec:SciFi}

Tidal forces are not the only danger a falling observer would face on the way to the black hole. They would obviously need to avoid the extremely hot accretion disk that inevitably orbits each black hole in the equatorial plane, as well as any jets that may escape from the poles. The trajectory of the fall would therefore need to be carefully chosen to avoid these hazards, while not straying too far from the axis of rotation. Moreover, as the observer gains speed during their fall, they would need to take care to minimize the risk of collision with particles that might be present in their immediate surroundings. Hawking radiation would pose no problem, as it is extraordinarily weak for very massive black holes. If the energy discontinuity at the event horizon (firewall) \cite{Almheiri:2012rt} assumed in certain models does not exist, then our observer will be able to penetrate below the event horizons. 
\\
The interior of a Kerr black hole remains unknown to us. It is therefore difficult to imagine just how dangerous the environment below the event horizons might be for an observer in free fall. However, assuming that the observer could eventually reach the center of the ring-shaped singularity, what would happen to them a few moments later? If we assume that the Kerr-Penrose diagram represents a physical reality, then we must accept that an observer could pass into another universe. However, they would keep all their observations to themselves, being unable to communicate them to the outside world by any means whatsoever. 
\\
If this scenario has even the slightest chance of corresponding to physical reality, then Kerr black holes may represent an escape route for any form of intelligence that might exist in the distant future, at a time when our universe would no longer be capable of sustaining any form of consciousness—whether due to a Big Crunch or to a Big Freeze. Let’s hope the new universe will be as welcoming as possible.

\section{Conclusion}
\label{sec:Conclu}

Taking into account only the gravitational effects, our computations show that the destiny of a free falling observer toward a black hole is very different if it is a Schwarzschild one or a Kerr one: unavoidable death for the former case, possibility of survival for the latter one, depending on the characteristics of the black hole. A trajectory along the rotation axis minimizes the tidal effect, and the ring-shaped singularity prevents the final destruction at the center of the black hole. 
If the Kerr black hole is massive enough and spins fast enough, the stress induced by the tidal effects can be lower than the maximal constraint bearable by a human body. 
This is probably the case for most of the supermassive black holes at the center of many galaxies, but not the case for stellar-mass astrophysical black holes even if their spin is non negligible. An analytical formula is obtained for the maximal mass of the black hole to allow survival. 
\\
As mentioned above, gravitational tidal forces are not the only hazards facing a fearless observer. The surroundings of black holes are extremely hostile, not to mention the unknown hazards hiding below the horizons. But the analysis of all these dangers is outside the scope of this paper. 


\bibliography{refs}

\end{document}